# Characterizing the VHE emission of LS I +61 303 using VERITAS observations

___________________________________________________________________

**D.B Kieda*[1] for the VERITAS Collaboration[†]**

[1]*Department of Physics and Astronomy University of Utah, Salt Lake City, UT 84112, USA*
*e-mail:* `dave.kieda@utah.edu`

The TeV gamma-ray binary LS I +61 303, approximately 2 kpc from Earth, consists of a low mass compact object in an eccentric orbit around a massive Be star. LS I +61 303 exhibits modulated VHE gamma-ray emission around its 26.5 days orbit, with strongest TeV emission during its apastron passage (orbital phases φ=0.55−0.65). Multiple flaring episodes with nightly flux variability at TeV energies have been observed since its detection in 2006. GeV, X-ray, and radio emission have been detected along the entire orbit, enabling detailed study of the orbital modulation pattern and its super-orbital period. Previously reported TeV baseline emission and spectral variations may indicate a neutron star flip-flop scenario, in which the binary system switches between accretor and propeller phases at different phases of the orbit.

Since September 2007, VERITAS has observed LS I +61 303 over three additional seasons, accruing 220+ hours of data during different parts of its orbit. In this work, we present a summary of recent and long-term VERITAS observations of LS I +61 303. This analysis includes a discussion of the observed variation of TeV emission during different phases of the orbit, and during different superorbital phases.

___________________________________________

*Speaker

[†] http://veritas.sao.arizona.edu; for collaboration list see PoS(ICRC2019)1177.

**Introduction**

The High Mass X-Ray Binary (HMXB) LS I +61∘ 303 was discovered by COS B in 1977 as an unidentified Galactic MeV gamma-ray emitter (2CG 135 +01) [1]. 2CG 135 +01 was found to be positionally coincident with variable radio source GT 0236 +610, which exhibited strong non-thermal radio bursts [2]. Long term radio observations of GT 0236 exhibited a 26.496 day (orbital) periodicity with a possible 4 year-long (superorbital) modulation of the radio amplitude [3]. The COS B data failed to show the reported 26.496 day radio periodicity. Subsequently, observations of 2CG 135 +01 by COMPTEL between 1991 and 1994 revealed a persistent hard gamma-ray source, but with a factor of 10-100 reduced flux compared to the COS B detection [4]. COMPTEL did not find evidence for the 26.5 day modulation of the >1 MeV gamma-ray signal, confirming the lack of modulation observed by COS B. GT 0236 +610 was identified by VLBI observations in 1993 as a binary system with two components separated by $3.1 \times 10^{11}$ (D/2.3 kpc) m [5], classifying it as a binary system. GT 0236 +610 was subsequently identified as associated with the optical source LS I +61º 303, which was spectroscopically identified as a massive (10-15 solar mass) early Be star [6].

LS I +61º 303 was also detected in X-rays [7, 8], exhibiting a hard X-ray spectrum consistent with other HMXB. LS I +61º 303 was first identified as a periodic TeV emitter by the MAGIC collaboration in 2006 [9], and subsequent observations of LS I +61º 303 by VERITAS demonstrated a combination of episodic flaring [10] with orbitally modulated (26.496 ± 0.0028 day period) TeV emission [11]. The strongest TeV emission is consistently between phase range φ = 0.55 − 0.65 near apastron (φ = 0.7775). Longer-term modulation of the TeV emission, consistent with superorbital modulation (1667 ± 8 day period) has also been demonstrated in a joint analysis of MAGIC and VERITAS observations [12]. In the GeV gamma-ray energies, LS I +61º 303 demonstrated orbitally modulation with a 6 GeV cutoff and strongest emission around φ = 0.3, as reported by Fermi observations [13]. The strongest GeV emission is coincident with the periastron passage of the compact object around the Be star. Both X-ray [14] and GeV emission (E < 5GeV) are found to be superorbitally modulated, with a dip in the periastron GeV maximum near superorbital phase 0.4, and a peak in the apastron GeV emission near superorbital phase 0.2 [15].

Previous multiwavelength observations of LS I +61º 303, using TeV observations by VERITAS and MAGIC combined with GeV observation by Fermi, and X-ray observation by SWIFT, have revealed conclusive evidence of a correlation between X-ray and TeV emission during flaring states [16,17,18], although earlier VERITAS studies failed to establish a significant correlation [19] . This correlation has been interpreted as requiring a common origin for both the X-Ray and TeV emission, providing evidence for an SSC or EC production mechanism. The multiwavelength observations also demonstrated an absence of any significant correlation between TeV emission and GeV emission. The lack of a correlation between GeV emission and X-ray/TeV emission has been interpreted as requiring a second, independent emission mechanism for the GeV emission.

An astrophysical model of LS I +61∘ 303 continues to emerge based upon the above (and other) multiwavelength observations [20]. The current model envisions a rapidly spinning Be star, spectral type B0Ve in a binary association with a compact object in an eccentric orbit (e = 0.54 ± 0.03) around a giant Be type 10-15 solar mass star. The Be star is thought to possesses a strong wind and a circumstellar decretion disk which crosses the plane of the orbit in two places, near orbital periastron [21]. The mass of the compact object is not completely determined, and arguments have been made for either a neutron star (NS) or a black hole with precessing jets. TeV gamma rays are produced by either the shock front of the colliding pulsar and Be star's winds, or by the relativistic precessing jets which are powered by accreting matter from the Be star onto the blackhole. More comprehensive models, such as the Ejector-Propeller model [22], which explores the effects of different radiative and wind pressures between apastron and periastron to explain the observed differences between the orbital GeV and TeV modulations. However, no current model adequately explains all the observed multiwavelength features of this system.

**The VERITAS IACT Observatory**

The VERITAS Observatory [23] is an array of 4 Imaging Atmospheric Cherenkov Telescopes (IACTs) located at the F.L. Whipple Observatory, Arizona, USA (31º 40N, 110º 57W, 11268 m.a.s.l). Each telescope features an f/1.0 12 m-diameter Davies-Cotton reflector consisting of 345 hexagonal facets, resulting in 110 $m^2$ of light collecting area. The separation between nearest neighbor telescopes is currently 80-120 m. The Davis-Cotton reflector provides a 4 nanosecond spread to photons arriving at the focal plane. The camera is pixelated with 499 photo-multiplier tubes (PMTs) each of 28.6 mm diameter. The PMTs are UV sensitive and have fast rise times coupled with high quantum efficiency. VERITAS has been upgraded twice: In 2009, Telescope 1 (T1) was relocated to a more symmetrical position/longer baseline in the array to increase sensitivity by 30% [24]. The second upgrade (2012) upgraded the photomultiplier tubes (PMTs) in each camera to increase average peak quantum efficiency from 20% at 320 nm to >32% at 330 nm [25]. This second upgrade also upgraded the Level 2 (camera level pixel trigger pattern) triggering system to improve trigger timing alignment, reduce L2 time coincident window, and increases rejection of random PMT triggered events [26]. The net effect of these two upgrades reduced VERITAS's gamma-ray threshold from 120 GeV down to 85 GeV, and an increase in VERITAS gamma-ray detection rate by 50% [27].

In its current configuration, VERITAS is sensitive to VHE gamma rays with energies between 85 GeV to > 30 TeV [28]. VERITAS nominally detects a source with 1% flux of the Crab Nebula in less than 25 h of observation.

**VERITAS LSI +61 303 Observations (2007-2016)**

VERITAS started observing LS I +61º 303 in 2006 and has continued its campaign every year, accumulating more than 200 h of observations up to December 2016. Previous LS I +61º 303 results used data collected between October 5, 2007 and November 23, 2016 covering three different epochs of the telescope operations [29]. Observations were performed in wobble mode with 0.5º offset, a standard mode for VERITAS observations. The combination of VERITAS latitude and source declination causes LS I+61º 303 to transit the sky at medium elevations; observations are within in 29º– 42º zenith. The entire dataset covers all orbital phases, with maximum exposures around apastron. The observations are quality-selected to remove minutes affected due to weather and hardware problems. Standard analysis for VERITAS point sources, using Event Display and VEGAS analysis packages, were performed on the data. Gamma-ray selection techniques used included using both box cuts and Boosted Decision Tree (BDT) algorithms [30]. After applying standard gamma-selection cuts, the observations have an effective gamma-ray threshold $E_{thresh}$ > 300GeV.

Using the zero orbital phase $\varphi_0 = MJD$ 43366.775 and an orbital period $P_{orb}$=26.4960 days, the data set is divided into 10 phases with width $\Delta\varphi = 0.1$. Except for three individual seasons (2008/2009, 2009/2010, and 2010/2011), the source is detected every season with significance varying between 5.6$\sigma$ and 21$\sigma$, generally near its apastron passage of the orbit. Observations performed during these ten years also demonstrate evidence for low-level persistent TeV emission (~3% Crab flux) [29].

TeV outbursts are generally observed from LS I +61o 303 close to its apastron passage in the phase range φ = 0.6 − 0.7 when flux observed can be well above 15% of the Crab Nebula. Photons with energy of 10 TeV or more have been observed during these flaring states implying the existence of energetic particles with tens of TeV energy in the binary system. During this period, there was a single episode of significant emission (5.6σ) from LS I +61º 303 during its superior conjunction, closer to its periastron passage [10].

**VERITAS LSI +61 303 Observations (2016-2019)**

During 2016-2019, VERITAS recorded an additional 32.7 hours of observations on LS I +61∘ 303 after weather and data quality cuts. The exposures in different years are distributed unevenly in various phase bins due to the full moon and other observing constraints. During 2016-2017 and 2018-2019, VERITAS observations primarily covered phases 0.4 to 0.9. These phases are near apastron, where the greatest TeV emission is traditionally observed. Observations in 2017-2018 covered phases near periastron (0.0 to 0.4) where TeV emission has generally been suppressed (Figure 1).

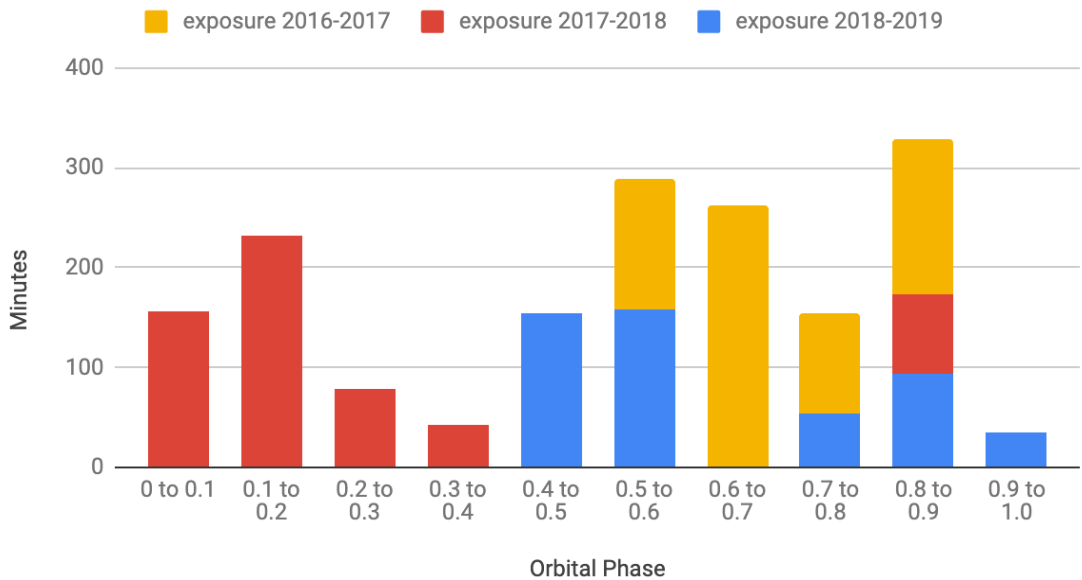

*Figure 1: VERITAS exposure on LS I +61 303 during 2016-2019, broken down according to orbital phase and histogrammed according to individual years. Yellow: 2016-2017 exposure. Red: 2017-2018 exposure. Blue: 2018-2019 exposure.*

Figure 2 shows the combined exposure of VERITAS LS I +61º 303 observations during the period 2007-2019, which spans three observing seasons (2016-2017, 2017-2018, and 2018-2019). During this period, more than 220 hours of data are available after weather and data quality cuts. Table 1 provides a detailed summary of the analysis results for each observing season. The table lists of the MJD range of the observations, the observed statistical significance $\sigma$ and normalized statistical significance $\sigma/\sqrt{hours}$, and the range of orbital phases and superorbital phases observed during each observing season. When integrating over each observing season, we observe a statistically strong detection (11.37 $\sigma$) of LS I +61º 303 only during 2016-2017, when both orbital phase range is near apastron. During this season, the superorbital phase (0.5) is also considered to be favorable, if previously report superorbital modulation trend [14] continues unabated. During 2017-2018 we observe only a marginal excess (2.72 $\sigma$) for the seasonal statistical significance as expected during LS I +61º 303's periastron passage. We also note that the superorbital phase of this season (0.8) predicts suppressed TeV emission.

For observing season 2018-2019, only a marginal excess (3.48 $\sigma$) is observed for LS I +61º 303 even though the observations were made around apastron, identical to 2016-2017. The 2018-2019 LS I +61º 303 observations were made at superorbital phase 0.0, which is predicted to coincide with the minimum TeV emission due to superorbital modulation. The seasonal VERITAS observations of LS I +61º 303 during the past three years are consistent with a continuation of the previously reported superorbital modulation of TeV emission[14].

At the conference, we expect to show an updated LS I +61º 303 analysis from the combined period 2007-2019.

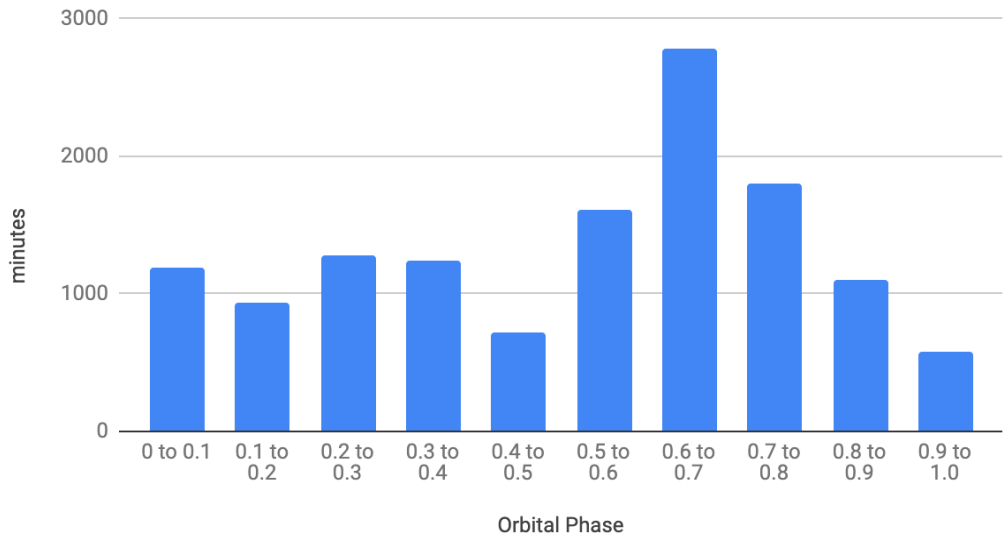

*Figure 2: Cumulative VERITAS exposure on LS I +61 303 during 2007-2019, broken down according to the orbital phase.*

*Table 1: Summary of VERITAS seasonal observations of LS I +61º 303. The right two columns indicate the orbital and supraorbital phase ranges for the corresponding observation date range. Green cells: Observations are within favorable phase ranges for strong emission. Red cells: Observations are within unfavorable phase range for strong emission. Observations within one or more unfavorable phase ranges are expected to have suppressed TeV gamma-ray emission.*

| Observing Season | MJD Observation Date Range | Total Significance $\sigma$ | Normalized. Significance $\sigma/\sqrt{hours}$ | Orbital Phase Range | Superorbital Phase |
|---|---|---|---|---|---|
| 2016-2017 | 57662-57696 | 11.37 | 3.32 | 0.5-0.8 | 0.5 |
| 2017-2018 | 58028-58051 | 2.72 | 0.77 | 0.8-1.4 | 0.8 |
| 2018-2019 | 58402-58492 | 3.48 | 1.18 | 0.4-0.9 | 0.0 |

## Acknowledgments

This research is supported by grants from the U.S. Department of Energy Office of Science, the U.S. National Science Foundation and the Smithsonian Institution, and by NSERC in Canada. This research used resources provided by the Open Science Grid, which is supported by the National Science Foundation and the U.S. Department of Energy's Office of Science, and resources of the National Energy Research Scientific Computing Center (NERSC), a U.S. Department of Energy Office of Science User Facility operated under Contract No. DE-AC02-05CH11231. We acknowledge the excellent work of the technical support staff at the Fred Lawrence Whipple Observatory and at the collaborating institutions in the construction and operation of the instrument.